\documentclass[a4paper,twocolumn,11pt,unpublished,]{quantumarticle}
\pdfoutput=1
\usepackage{amsmath}
\usepackage{amscd}
\usepackage{wasysym}
\usepackage{hyperref}
\usepackage{color}
\usepackage{braket}
\usepackage{float}
\usepackage{lipsum}
\usepackage[numbers,sort&compress]{natbib}

\usepackage[T1]{fontenc}
\usepackage{ae,aecompl}
\usepackage{amsfonts}
\usepackage{amsthm}
\usepackage{dsfont}
\usepackage{graphicx}
\usepackage{subcaption}
\usepackage{amsfonts}
\usepackage{physics}

\newcommand{\ue}{\mathrm{e}}
\newcommand{\dr}{\mathrm{d}}

\renewcommand\Re{\operatorname{Re}}

\newcommand{\la}{\langle}
\newcommand{\ra}{\rangle}

\hypersetup{colorlinks=true,
    citecolor=blue,
    linkcolor=red,
    urlcolor=blue}

\begin{document}

\title{Steady-state Bell nonlocality in an autonomous quantum thermal machine}

\author{Bradley Longstaff}
\affiliation{Department of Physics and Astronomy, University of Exeter, Exeter, EX4 4QL, United Kingdom}
\orcid{0000-0002-0928-2505}

\maketitle

\begin{abstract}
  A quantum thermal machine is presented that is able to autonomously generate steady-state Bell nonlocality. A Lindblad equation is derived for two qubits that are incoherently coupled to a pair of thermal baths and the resulting Liouvillian is found to have a strong symmetry. This out-of-equilibrium system can generate Bell-nonlocal steady states across a range of parameters and at arbitrarily high temperatures for certain initial conditions. To analyse how the machine operates in a more realistic setting, experimental imperfections are then included via a stochastic perturbation to the system Hamiltonian. This breaks the strong symmetry and results in collective and local dephasing noise. It is found that if the sole source of noise is local noise, then the steady state is Bell local. The co-presence of collective noise is then shown to be advantageous, with the ability to increase the degree of Bell-inequality violation and transform a steady state from Bell local to Bell nonlocal.
\end{abstract}

\section{Introduction}
\label{sec.intro}

Entanglement and quantum correlations are key features of quantum physics and play an important role in many quantum-information-processing tasks \cite{Horodecki2009,Gisin07,giovanetti2011}. Bell demonstrated that there are quantum correlations that cannot be explained in terms of a locally-causal theory \cite{Einstein1935,Bell1964}, provided that certain assumptions hold \cite{Ney21,Adlam21}. Bell inequalities are derived from the constraints that locally-causal theories place on possible experimental observations and they can be violated by performing local measurements on a shared entangled state \cite{Hens15,Shal15,Gius15}. Bell-nonlocal states, the states that can violate a Bell inequality, are essential for highly-secure information-processing tasks such as device-independent quantum key distribution \cite{Pironio09} and random number generation \cite{colbeckPhD2009,Pironio2010}. 

Environment-assisted entanglement generation has been pursued as a technique to generate and stabilise entanglement \cite{GenEnt1,GenEnt2,GenEnt3,GenEnt4,GenEnt5,GenEnt6,GenEnt7}, including the use of dissipation and external driving to obtain entangled steady states \cite{SteadyState1,SteadyState2,SteadyState3,SteadyState4}. It is also possible to generate steady-state entanglement in an autonomous setting, where no source of coherence or external control is required \cite{NoExtForce1,NoExtForce2,NoExtForce3,NoExtForce4,NoExtForce5,NoExtForce6,NoExtForce7,Tavakoli2020}. 

Two-qubit thermal machines have attracted attention for their simplicity and capacity to generate steady states that can perform useful tasks such as steering, teleportation and Bell inequality violation \cite{Brask2015,Man2019,Brask2022,Khandelwal20}. In the simplest set ups considered so far, steady-state entanglement is possible but is typically weak and cannot be used to violate Bell inequalities \cite{Brask2022}. Additional resources are then needed to obtain Bell-nonlocal steady states. These can be generated, for example, in a heralded manner by supplementing the thermal machine with local filters \cite{Tavakoli2018,Brask2022}, or by implementing a continuous-feedback protocol \cite{Diotallevi24}.

This work shows that steady-state Bell nonlocality can be generated in a simple two-qubit autonomous thermal machine without having to introduce additional resources. Previous works have assumed the presence of a free-evolution term in the system Hamiltonian, arising from qubits with a fixed energy gap. Yet this term is not always present, for example, in Bose-Hubbard dimers and the Su-Schrieffer-Heeger model. This paper therefore starts by analysing a thermal machine without the free-evolution term and finds that the resulting open quantum system exhibits a strong symmetry \cite{Buca12}. This difference is significant because symmetries are known to help preserve coherence \cite{Dubois23}, stabilise long-range coherence in lossy qubit arrays \cite{Dutta20}, and assist in the preservation of entanglement \cite{Li25}. This work demonstrates that a strong symmetry can enable Bell-nonlocal steady states to be generated at arbitrarily high temperatures.

In practice, experimental imperfections can break a strong symmetry. For example, cold atoms trapped in a two-site optical lattice can be described by a Bose-Hubbard dimer \cite{Jaksch98}. In theory, the on-site energy (the `free-evolution' term) can be set to zero. In practice, however, the optical lattice will be subject to stochastic lattice noise, resulting in on-site energies that fluctuate around zero. This paper therefore considers experimental imperfections in the form of local and collective dephasing noise. The free-evolution terms therefore appear in this work as stochastic perturbations, in contrast to the static terms that have previously been considered.

It is then shown that the introduction of noise breaks the strong symmetry, maintains a weak symmetry, and results in a unique steady state that can again be Bell nonlocal. Remarkably, the machine can use collective dephasing noise to an advantage \cite{Plenio08,Feldmann06,Znidaric09}, with the ability to improve the degree of steady-state Bell nonlocality and transform a steady state from Bell local to Bell nonlocal. Natural units are assumed throughout the paper, with the Boltzmann constant $k_B = 1$ and $\hbar = 1$. The states $|\psi_\pm\rangle, |\phi_\pm\rangle$ are the usual two-qubit Bell states.

\section{The thermal machine}
\label{sec:ThermalMachine}

The machine consists of a  pair of interacting qubits, weakly coupled to two separate thermal baths. The Hamiltonian of the two-qubit system $\hat{H}_S$ generates a flip-flop interaction between the qubits with strength $g>0$,
\begin{equation}
    \label{eqn:sys ham}
    \hat{H}_S = g\left(|\psi_-\rangle\langle \psi_-| - |\psi_+\rangle \langle \psi_+|\right),
\end{equation}
and the environment is described by the non-interacting Hamiltonian
\begin{equation}
	\hat{H}_E = \sum_{\beta=1}^{2} \hat{H}_{E_\beta} \quad \textnormal{with} \quad \hat{H}_{E_\beta} = \sum_{q=0}^\infty \Omega_{\beta q} \, \hat{c}^\dag_{\beta q} \hat{c}_{\beta q}.
\end{equation}
The bath spectra are non negative $\Omega_{\beta q} \geq 0$ and each bath can consist of spins or bosons. For the bosonic baths, the bath creation and annihilation operators satisfy the canonical commutation relations
\begin{align}
	[\hat{c}_{\alpha k},\hat{c}^\dag_{\beta q}] &= \delta_{\alpha \beta}\delta_{kq},\\
	[\hat{c}_{\alpha k},\hat{c}_{\beta q}] &= [\hat{c}^\dag_{\alpha k},\hat{c}^\dag_{\beta q}] = 0,
\end{align}
while spins in different spin baths commute and spins within the same bath obey the anticommutation relations
\begin{align}
\{\hat{c}_{\beta k},\hat{c}^\dag_{\beta q}\} &= \delta_{kq},\\
	\{\hat{c}_{\beta k},\hat{c}_{\beta q}\} &= \{\hat{c}^\dag_{\beta k},\hat{c}^\dag_{\beta q}\} = 0.
\end{align}
Let the system-bath interaction be of the form
\begin{equation}
     \label{eqn:s-b op}
    \hat{H}_{SE} = \sum_{\beta = 1}^2 \hat{A}_\beta \otimes \hat{B}_\beta,
\end{equation}
where $\hat{A}_\beta = \hat{\sigma}_{x \beta}$ is the Pauli spin-$x$ operator for qubit $\beta$ and
\begin{equation}
    \hat{B}_\beta = \sum_{q=0}^\infty g_{\beta q}\left(\hat{c}^\dag_{\beta q} + \hat{c}_{\beta q}\right)
\end{equation}
acts on the Hilbert space of the environment. The real-valued coefficients $g_{\beta q}$ quantify the coupling strength of each qubit to its respective bath. 

The system Hamiltonian \eqref{eqn:sys ham} does not contain any free-evolution terms. This could be realised, for example, with cold atoms trapped in a double-well optical lattice. In this bosonic system, strong interactions would be used to implement the hard-core boson constraint, restricting the basis to $\{|11\rangle,|10\rangle,|01\rangle,|00\rangle\}$. The system-environment interaction \eqref{eqn:s-b op} then corresponds to coupling the position quadratures $\hat{A}_\beta = \hat{a}^\dag_\beta + \hat{a}_\beta$ to the baths, where $\hat{a}^\dag_\beta,\hat{a}_\beta$ are bosonic creation and annihilation operators for mode $\beta$. Under the hard-core boson constraint this can effectively be written as $\hat{A}_\beta = \hat{\sigma}_{x \beta}$.

If the environment is initially in the thermal state
\begin{equation}
	\label{eqn:rhoE}
\hat{\rho}_E = \frac{\ue^{-\hat{H}_{E_1}}}{\tr\left[\ue^{-\hat{H}_{E_1}}\right]} \otimes \frac{\ue^{-\hat{H}_{E_2}}}{\tr\left[\ue^{-\hat{H}_{E_2}}\right]}
\end{equation}
then the two-point bath correlation functions reduce to 
\begin{align}
	\la \hat{c}^\dag_{\alpha k}\hat{c}_{\beta q}\ra = p_\beta(\Omega_{\beta q}) \, \delta_{\alpha \beta} \, \delta_{kq},\\
	\la \hat{c}_{\alpha k}\hat{c}_{\beta q}\ra = 0, \quad \la \hat{c}^\dag_{\alpha k}\hat{c}^\dag_{\beta q}\ra = 0,
\end{align}
where the expectation values $\la \cdot \ra$ are taken in the state $\hat{\rho}_E$ and the distribution
\begin{equation}
    \label{eqn:prob dists}
	p_\beta(\Omega) = \left[\ue^{\Omega/T_\beta}+\xi_\beta\right]^{-1}
\end{equation}
is the Bose-Einstein distribution for a bosonic bath ($\xi_\beta = -1$) at temperature $T_\beta$, or the Fermi-Dirac distribution for a spin bath ($\xi_\beta = 1$).

As the system is weakly coupled to the environment, a Lindblad master equation can be derived to approximate the system dynamics. Several conditions must hold for this to be possible (as full details can be found elsewhere \cite{Rivas12,Rivas2010}, only the key ideas are noted here).  A Lindblad equation can be derived provided that: the system-bath interactions are weak; the initial system-bath state factorises and the environment is in an equilibrium state $[\hat{H}_E,\hat{\rho}_E]=0$; the correlations in the baths decay much faster than variations of the system state in the interaction picture; and the secular approximation is valid, placing constraints on the Bohr frequencies of the system Hamiltonian $\hat{H}_S$. 

For the thermal machine outlined above, the Lindblad equation for the reduced state $\hat{\rho}$ of the two-qubit system has the form
\begin{align}
    \label{eqn:lindblad eqn}
    \frac{\dr \hat{\rho}(t)}{\dr t} &= -i\left[\hat{H}_S + \hat{H}_{LS},\hat{\rho}(t) \right] \nonumber\\ & +  \sum_{\omega} \sum_{\beta} \gamma_{\beta}(\omega) \mathcal{D}\left(\hat{A}_{\beta}(\omega)\right)[\hat{\rho}(t)],
\end{align}
where, given an operator $\hat{Y}$, the superoperator $\mathcal{D}(\hat{Y})$ acts on a quantum state $\hat{\rho}$ as
\begin{equation}
    \label{eqn:dissipator op}
    \mathcal{D}(\hat{Y})[\hat{\rho}] = 
     \hat{Y} \hat{\rho}\hat{Y}^\dag - \frac{1}{2} \left\{\hat{Y}^\dag \hat{Y},\hat{\rho}\right\}.
\end{equation}
The right-hand side of the Lindblad equation \eqref{eqn:lindblad eqn} can be viewed as a linear superoperator (the Liouvillian) acting on the state, $\mathcal{L}[\hat{\rho}(t)]$.

Only two Bohr frequencies $\omega \in \{g,-g\}$ of the system Hamiltonian contribute to the non-unitary dynamics, with the four jump operators
\begin{align}
\label{eqn:jump ops}
\hat{A}_1(g) &= |\psi_+\rangle\langle \phi_+|-|\phi_-\rangle\langle \psi_-|, \\
\hat{A}_2(g) &= |\psi_+\rangle\langle \phi_+|+|\phi_-\rangle\langle \psi_-|,
\end{align}
and $\hat{A}_\beta(-\omega)=\hat{A}^\dag_\beta(\omega)$ (see Appendix \ref{app:eig and jump}). The real-valued and positive rates are
\begin{equation}
    \label{eqn:gamma rates}
    \gamma_\beta(\omega) = \begin{cases} 
J_\beta(\omega)\left[1-\xi_\beta P_\beta(\omega)\right] & \textrm{if}\quad \omega > 0, \\ 
J_\beta(|\omega|)P_\beta(|\omega|) & \textrm{if} \quad \omega<0, 
\end{cases}
\end{equation}
where $J_\beta(\omega)$ is the spectral density of bath $\beta$.

To be consistent with the secular approximation, the inter-qubit coupling strength must be much larger than the decay rates, $g \gg \max_{\beta ,\omega}\{\gamma_\beta(\omega)\}$. The Lindblad equation also requires that the Bohr frequencies of the system are much smaller than the cut-off frequencies appearing in the spectral-density functions. Therefore, when it is necessary to specify a form for $J_\beta(\omega)$, a linear frequency dependence $J_\beta(\omega) = \chi_\beta \omega$ will be assumed, where $\chi_\beta$ is a positive coupling parameter.

Interaction with the environment also results in a unitary contribution to the dynamics given by the Lamb shift $\hat{H}_{LS} = \sum_\omega \sum_\beta S_\beta(\omega) \hat{A}^\dag_\beta(\omega)\hat{A}_\beta(\omega)$. The approximations used to derive the Lindblad equation require this term to be small and it is often dropped. In the model presented here it is found to have no influence on the steady state, regardless of the values of $S_\beta(\omega)$, so it does not need to be neglected.

\section{Strong symmetry and steady states}
To determine the strong symmetry of the Liouvillian, first consider the two-qubit spin-$x$ parity operator $\hat{X} = \hat{\sigma}_x\otimes\hat{\sigma}_x$ and how it acts on the Bell states
\begin{equation}
    \label{eqn:bell symmetry}
    \hat{X}|\phi_\pm\rangle = \pm|\phi_\pm\rangle, \quad \hat{X}|\psi_\pm\rangle = \pm|\psi_\pm\rangle.
\end{equation}
Observe that $\hat{X}$ commutes with the system Hamiltonian and the Lamb-shift Hamiltonian $[\hat{X},\hat{H}_S] = 0 = [\hat{X},\hat{H}_{LS}]$, as well as with every jump operator $[\hat{X},\hat{A}_\beta(\omega)]=0$. Spin-$x$ parity is therefore a strong symmetry and the Liouvillian $\mathcal{L}$ decouples into two symmetry sectors, $\mathcal{L} = \mathcal{L}_\textrm{even} \oplus \mathcal{L}_\textrm{odd}$ \cite{Buca12,Albert2014}. The even-parity subspace consists of operators that are invariant under $\hat{X}$, and the odd-parity subspace contains those that pick up a minus sign. The even subspace further decomposes into $\mathcal{L}_\textrm{even} = \mathcal{L}_{+} \oplus \mathcal{L}_{-}$, where the $4 \times 4$ blocks $\mathcal{L}_\pm$ are spanned by all the outer products of $\{|\phi_\pm\rangle,|\psi_\pm\rangle\}$ respectively. The odd block is spanned by the traceless cross operators composed of opposite-parity Bell states, e.g., $|\psi_+\rangle\langle\phi_-|$, and no physical state can be wholly contained in this block.

The right and left eigenoperators of the Liouvillian are needed to work out the steady states, where the right $\hat{r}_j$ and left $\hat{l}_j$ eigenoperators corresponding the to eigenvalue $\lambda_j$ satisfy $\mathcal{L}[\hat{r}_j] = \lambda_j \hat{r}_j$ and $\mathcal{L}^\dag[\hat{l}_j] = \lambda^\ast_j \hat{l}_j$. The steady states are the right eigenoperators with eigenvalue zero and each sector $\mathcal{L}_\pm$ contains at least one physical steady state. By writing the Liouvillian in the Bell basis it is straightforward to find  two steady states, one contained in each sector and denoted $\hat{r}_+$ and $\hat{r}_-$ respectively,
\begin{align}
    \label{eqn:right eig ops}
    \hat{r}_+ &=  a|\psi_+\rangle\langle \psi_+| + (1-a)|\phi_+\rangle\langle \phi_+|,\\
    \hat{r}_- &= a|\phi_-\rangle\langle \phi_-| + (1-a)|\psi_-\rangle\langle \psi_-|.
\end{align}
Each steady state in this pair is a convex combination of Bell states with $a = \Gamma_1/(\Gamma_1+\Gamma_2)$ and
\begin{equation}
\label{eqn:big gammas}
    \Gamma_1 = \gamma_1(g)+\gamma_2(g), \quad \Gamma_2 = \gamma_1(-g)+\gamma_2(-g).
\end{equation}
The corresponding left eigenoperators are the projectors onto the plus and minus subspaces, $\hat{l}_+ = \hat{\Pi}_+$ and $\hat{l}_- = \hat{\Pi}_-$. To see this, observe that $\mathcal{L}^\dag[\hat{X}] = 0$ and $\mathcal{L}^\dag[\hat{I}] = 0$. Then by linearity of $\mathcal{L}^\dag$, the projectors $\hat{\Pi}_+ = (\hat{I}+\hat{X})/2$ and $\hat{\Pi}_- = (\hat{I}-\hat{X})/2$ are left eigenoperators with eigenvalue zero. The remaining left and right eigenoperators, corresponding to the non-zero eigenvalues, do not need to be calculated. Note, however, that the complete set forms a biorthogonal basis $\tr[\hat{l}^\dag_j\hat{r}_k] = \delta_{jk}$. Expanding the initial state $\hat{\rho}(0)$ in this basis and applying the dynamical map gives
\begin{equation}
    \hat{\rho}(t) = \ue^{t \mathcal{L}}[\hat\rho(0)] = \sum_j \tr\left[\hat{l}^\dag_j \hat{\rho}(0)\right]\hat{r}_j\ue^{\lambda_j t}.
\end{equation}
Because the non-zero eigenvalues all have negative real parts $\Re[\lambda_j] < 0$ the steady states have the form
\begin{equation}
    \label{eqn:steady state final}
    \hat{\rho}_\infty = \mu\hat{r}_+ + (1-\mu)\hat{r}_-,
\end{equation}
 and the weight of the initial state in the plus sector $\mu = \tr[\hat{\Pi}_+\hat{\rho}(0)]$ determines the steady state.
 
\section{Noiseless systems}
First consider a perfectly operating thermal machine in which the strong spin-$x$ parity symmetry is unbroken. This will determine how Bell nonlocality can vary with temperature in principle. The effects of experimental noise that breaks the strong symmetry will be analysed later.

The simplest Bell scenario consists of two parties, each performing measurements with binary inputs and outputs, and is completely characterised by the CHSH inequality \cite{Clauser69}. Given an arbitrary two-qubit state $\hat\rho$, the Horodecki criterion provides the maximum attainable CHSH value under optimal measurements \cite{Horodecki95}. This maximum is given by $\mathcal{B} = 2\sqrt{\lambda_1 + \lambda_2}$, where $\lambda_1$ and $\lambda_2$ are the two largest eigenvalues of the $3\times 3$ matrix $T^T T$, where $T_{ij} = \tr[\hat\sigma_i \otimes\hat\sigma_j \hat\rho]$ and $\hat\sigma_j$ are the Pauli operators. If the Bell parameter $\mathcal{B} > 2$, then the state $\hat\rho$ will violate the CHSH inequality for some set of measurements. The largest violation possible with quantum theory is $\mathcal{B} = 2\sqrt{2}$ and this bound can be reached with a maximally entangled state \cite{Cirelson80}.

Applying the Horodecki criterion to the steady state \eqref{eqn:steady state final} yields the Bell parameter 
\begin{equation}
    \label{eqn:chsh 1}
    \mathcal{B} = 2\sqrt{\left(\frac{\Gamma_1-\Gamma_2}{\Gamma_1+\Gamma_2}\right)^2+(2\mu-1)^2}.
\end{equation}
It is immediately clear that initial states with support entirely on the plus or minus sectors $\mu\in\{0,1\}$, evolve to Bell-nonlocal steady states whenever $\Gamma_1 \neq \Gamma_2$, which is true for all finite temperatures \eqref{eqn:big gammas}. Two qubits initialised in spin-$x$ up eigenstates can exhibit this steady-state behaviour. On the other hand, initial states with $\mu = 1/2$ always tend to a Bell-local steady state, such as the initial states $|\psi(0)\rangle \in \{|00\rangle,|01\rangle,|10\rangle,|11\rangle$\}. Whenever $\mu \neq 1/2$, there is a pair of bath temperatures that will result in a Bell-nonlocal steady state.

In order to gain further insight into how the Bell nonlocality varies with temperature and $\mu$, consider the case of equal temperature baths, where the Bell parameter simplifies to
\begin{equation}
    \label{eqn:chsh thermal}
    \mathcal{B} = 2\sqrt{\tanh^2\left(\frac{g}{2T}\right)+(2\mu-1)^2}.
\end{equation}
The CHSH inequality is violated whenever $|2\mu-1| > \sech(g/2T)$. For small values of $T/g$, a wide range of $\mu$ can give $\mathcal{B}>2$. Increasing $T/g $ shrinks the hyperbolic tangent in \eqref{eqn:chsh thermal}, restricting the values of $\mu$ that allow for a Bell-inequality violation. As $\sech(g/2T)$ approaches unity for large $T/g$, only initial states with $\mu\in\{0,1\}$ will yield Bell-nonlocal steady states for arbitrarily large $T/g$.

As the bath temperatures must be consistent with the approximations of the master equation, the phrase `arbitrarily large' needs to be clarified. There are no temperature restrictions on spin baths because they saturate at high temperatures. It therefore follows that Bell nonlocality can persist no matter how high the temperature. However, the Bose-Einstein distribution grows linearly with $T/g$ at high temperatures and the consistency condition $g \gg \max_{\beta,\omega}\{\gamma_\beta(\omega)\}$ requires that  $1/\chi_\beta \gg T_\beta/g$ for bosonic baths. If at least one bath is bosonic, the results above demonstrate the possibility of Bell-nonlocal steady states for all temperatures satisfying this upper bound.

\begin{figure}[h]
    \centering
		\includegraphics[width=0.48\textwidth]{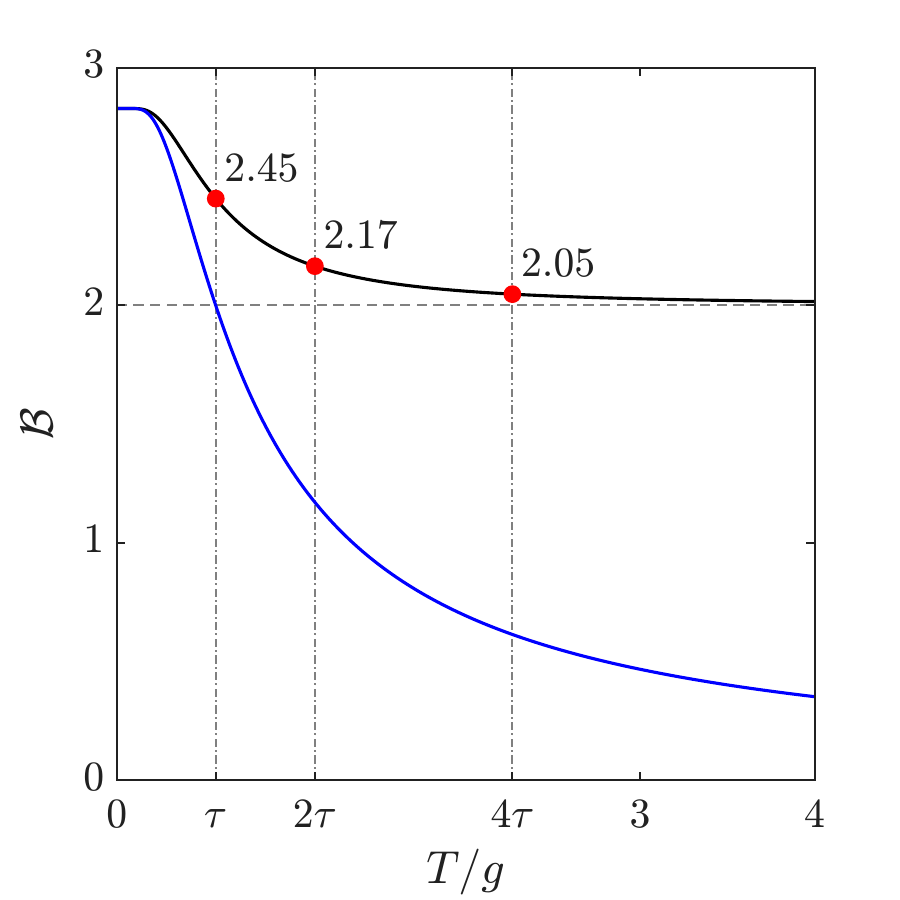}
	\caption{Maximum attainable CHSH
value $\mathcal{B}$ as a function of the ratio of temperature $T$ to the inter-qubit coupling strength $g$, for a non-equilibrium steady state (NESS) with $\mu=1$ (black curve) and the thermal state (blue curve). The thermal state becomes Bell local at the critical temperature $ T = \tau g$, while the NESS is Bell nonlocal for all $T/g$. The value of $\mathcal{B}$ is labelled for several multiples of $\tau$.}
    \label{fig:NEvsTH}
\end{figure}

When the ratio $T/g$ is small enough the thermal state
\begin{equation}
\label{eqn:thermal state}
    \hat{\rho}_\textrm{th} = \ue^{-\hat{H}_S/T}/\tr[\ue^{-\hat{H}_S/T}]
\end{equation}
can violate a CHSH inequality (see Appendix \ref{app:thermal}), which raises the question of how the non-equilibrium steady states (NESSs) of the thermal machine \eqref{eqn:steady state final} compare to the thermal state. Suppose that the thermal state violates the CHSH inequality at $T/g$. The Bell parameter \eqref{eqn:chsh thermal} then indicates that any initial state with $|2\mu-1|>|2\mu_\textrm{th}-1|$ will evolve to a NESS that can more strongly violate the CHSH inequality, where $\mu_\textrm{th} = (\ue^{-g/T}+1)^{-1}$. At large $T/g$, the NESS behaviour can differ substantially from the thermal state. This is because the latter becomes Bell local at the critical temperature $T = \tau g$, with $\tau = 1/(2\ln[\sqrt{2}+1])$, and tends to the maximally mixed state as $T/g \to \infty$. Fig. \ref{fig:NEvsTH} illustrates a NESS that significantly violates the CHSH inequality at the critical temperature of the corresponding thermal state.
\begin{figure*}[t]
    \centering
    \hspace*{-0.9cm}  
        \includegraphics[width=\textwidth]{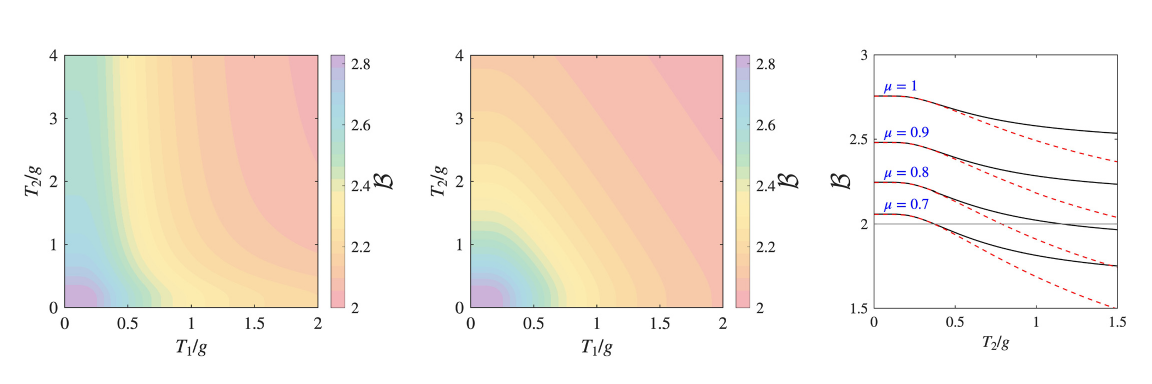}
	\caption{(Left) Spin baths saturate at high temperatures, allowing for persistent steady-state Bell nonlocality. The maximum attainable CHSH value $\mathcal{B}$ is plotted for a non-equilibrium steady state with $\mu=1$ and various values of $T/g$ for a pair of spin baths. The system is more strongly coupled to the bath 1, with $\chi_1 = 0.015$ and $\chi_2 = 0.005$. (Middle) The negative impacts of thermal noise on $\mathcal{B}$ for a pair of bosonic baths. Parameters are the same as in the left plot. (Right) Illustration of how this behaviour varies with the weight parameter $\mu$ (or initial state), when the temperature of bath 1 is fixed at $T_1 = 0.3g$. The black line corresponds to the spin baths and the red-dashed line corresponds to the bosonic baths.}
    \label{fig:T1vsT2}
\end{figure*}

It is not possible to boost the steady-state value of $\mathcal{B}$ obtained with equal-temperature baths by operating the thermal machine between a hot bath and a cold bath. That is, $\mathcal{B}(T_1,T_2) < \mathcal{B}(T_2,T_2)$ when $T_1 > T_2$. Increasing the temperature of one bath can only degrade the steady-state Bell nonlocality. This short calculation is presented in Appendix \ref{app:no boost}. Nevertheless, the steady-state Bell nonlocality can still be substantial when $T_1 \neq T_2$. This is particularly true if a spin bath is involved as it will saturate at high temperatures, allowing for persistent Bell nonlocality. In contrast, the decay rates of bosonic baths grow linearly with $T/g$ at high temperatures and the resulting thermal noise more rapidly decreases the maximum attainable CHSH value (Fig. \ref{fig:T1vsT2}).

\section{Experimental imperfections}
The previous section established that a strong symmetry enables the generation of NESSs that are Bell nonlocal at arbitrarily high $T/g$. Although it is interesting that quantum mechanics predicts Bell nonlocality at high temperatures, in practice there can be experimental noise in the implementation of $\hat{H}_S$ that breaks the strong symmetry. This can radically alter the steady-state dynamics of the thermal machine.

Uncontrolled environmental fluctuations can be modelled with a stochastic perturbation $\hat{V}(t)$ to the system Hamiltonian, $\hat{H}_S \to \hat{H}_S + \hat{V}(t)$. Dephasing is often the dominant form of noise and the qubits can be subjected to collective and local dephasing environments
\begin{equation}
\label{eqn:V pert}
    \hat{V}(t) = \begin{cases}
    \eta_c(t)\left(\hat{\sigma}_{z1} + \hat{\sigma}_{z2}\right), & \textrm{collective noise}\\
    \eta_{l_1}(t) \hat{\sigma}_{z1} + \eta_{l_2}(t) \hat{\sigma}_{z2},& \text{local noise }
\end{cases}
\end{equation}
where $\eta_j(t)$ are independent classical white noise processes with $\mathbb{E}[\eta_j(t)] = 0$ and $\mathbb{E}[\eta_j(t)\eta_k(t')] = \gamma_j\delta_{jk} \delta(t-t')$. While the collective noise with intensity $\gamma_c$ models shared environmental fluctuations, the local noise describes independent noise sources with intensities $\gamma_{l_1}$ and $\gamma_{l_2}$ that act on each qubit separately. The subscripts distinguish these quantities from the decay rates \eqref{eqn:gamma rates}. Let $\gamma_l = (\gamma_{l_1}+\gamma_{l_2})/2$ in what follows.

The inclusion of noise leads to a stochastic differential equation for the total system-environment state and taking the ensemble average yields a deterministic master equation for the ensemble-averaged total state (see, e.g., \cite{Kiely21}). The coupling to the two thermal baths can then be found using the methods outlined above, leading to a Liouvillian with three additional terms
\begin{equation}
\label{eqn:liouvillian perturbed}
\mathcal{L} = \mathcal{L}_0 + \gamma_c\mathcal{D}\left(\hat{\sigma}_{z1}+\hat{\sigma}_{z2}\right) + \gamma_{l_1}\mathcal{D}\left(\hat{\sigma}_{z1}\right)+\gamma_{l_2} \mathcal{D}\left(\hat{\sigma}_{z2}\right),
\end{equation}
where $\mathcal{D}$ was defined in \eqref{eqn:dissipator op} and $\mathcal{L}_0$ is the unperturbed Liouvillian \eqref{eqn:lindblad eqn}. In addition to the assumptions needed to derive $\mathcal{L}_0$, this master equation also demands that the bath correlations decay much faster than the classical noise-induced dephasing influences the interaction-picture system dynamics, $ \tau_B \ll 1/[4(2\gamma_c+\gamma_l)]$. 

Noise has broken the strong symmetry but the system still has a weak spin-$x$ parity symmetry, $\mathcal{L}[\hat{X}^\dag \hat\rho \hat{X}] = \hat{X}^\dag\mathcal{L}[\hat\rho]\hat{X}$ \cite{Albert2014,Buca12}. The Liouvillian thus block diagonalises into $\mathcal{L} =\mathcal{L}_\textrm{even} \oplus \mathcal{L}_\textrm{odd}$ and again the steady state must be contained in the even block. In this case, the system now has a unique steady state (see Appendix \ref{app:noisy results}).
 
Substituting the steady state into the Horodecki criterion yields the Bell parameter (see Appendix \ref{app:noisy results})
\begin{equation}
\label{eqn:horodecki noise}
    \mathcal{B} = 2\sqrt{\gamma_c^2+(\gamma_c+\gamma_l)^2}\left(\Gamma_1-\Gamma_2\right)/N,
\end{equation}
where $N = 4\gamma_l^2+8\gamma_c\gamma_l + (\gamma_c+\gamma_l)(\Gamma_1+\Gamma_2)$. This has three immediate properties. First, the derivation in Appendix \ref{app:no boost} can also be applied here to demonstrate that a temperature gradient can only degrade any steady-state Bell nonlocality, just like the unperturbed system. Second, if the only source of noise is local noise then the steady state is always Bell local. Finally, if only collective noise is present, the Bell parameter \eqref{eqn:horodecki noise} simplifies to $\mathcal{B} = 2\sqrt{2}\left(\Gamma_1-\Gamma_2\right)/\left(\Gamma_1+\Gamma_2\right)$ and the steady state can exhibit Bell nonlocality. Surprisingly, this expression is independent of the collective-noise intensity $\gamma_c$. This independence follows from the fact that, in this particular situation, the steady-state condition $\mathcal{L}[\hat{\rho}_\infty]=0$ also implies that $\mathcal{D}(\hat{\sigma}_{z1}+\hat{\sigma}_{z2})[\hat{\rho}_\infty] = 0$.
\begin{figure}[h]
    \centering
    \hspace*{-0.5cm}  
	\includegraphics[width=0.5\textwidth]{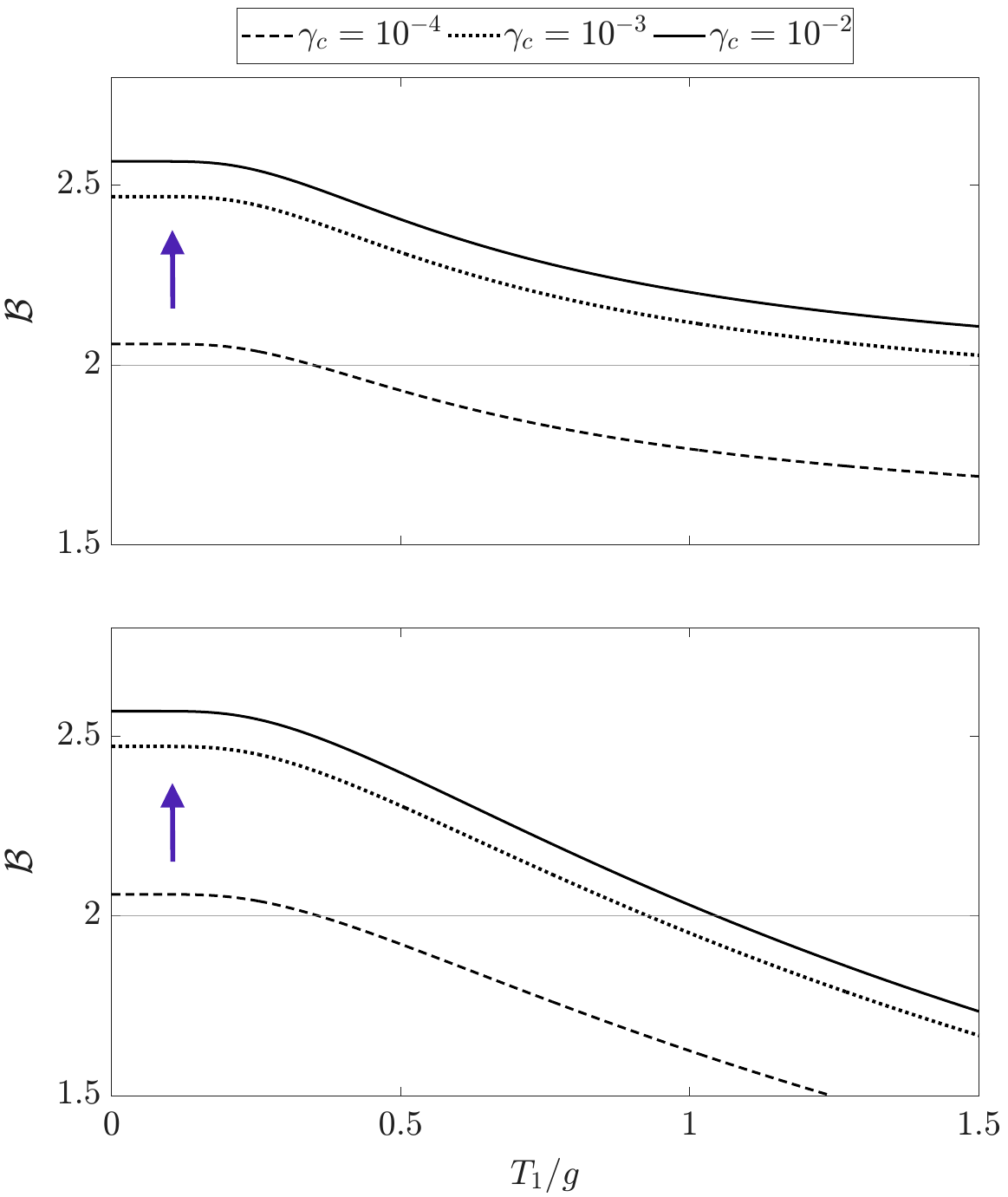}
	\caption{Increasing the collective noise intensity $\gamma_c$ can boost steady-state Bell nonlocality. The temperature of bath 2 is fixed at $T_2=0.3g$ and the Bell parameter $\mathcal{B}$ is plotted as a function of $T_1/g$, for two spin baths (top row) and two bosonic baths (bottom row). Each qubit experiences a local dephasing noise intensity of $\gamma_{l} = 10^{-4}$. The bath coupling parameters are $\chi_1 = 0.005$ and $\chi_2 = 0.015$. The arrow illustrates the direction of increasing collective noise.}
    \label{fig:noisy plot}
\end{figure}

These results suggest that the presence or lack of steady-state Bell nonlocality will strongly depend on the ratio $\gamma_l/\gamma_c$ when both types of noise are included. By fixing $\gamma_l$ and expressing the Bell parameter \eqref{eqn:horodecki noise} in terms of $\gamma_l/\gamma_c$ it can be shown that steady-state Bell nonlocality is possible when 
\begin{equation}
    \label{eqn:nonlocal cond noise}
    \Gamma_1 > (3+2\sqrt{2})\Gamma_2 + 8(1+\sqrt{2})\gamma_l
\end{equation}
(see Appendix \ref{app:noisy nl conds}). By substituting in $\Gamma_1$ and $\Gamma_2$ \eqref{eqn:big gammas}, it is found that the weak-coupling condition strongly constrains the values of $\gamma_l$ for which $\mathcal{B}>2$ is possible. However, \eqref{eqn:nonlocal cond noise} can be fulfilled provided that $\gamma_l$ is small enough.

Therefore, even though the steady state is Bell local when only local noise is present, Bell-nonlocal steady states can be generated by introducing collective noise (Fig. \ref{fig:noisy plot}). The machine turns collective noise into an advantage \cite{Plenio08,Feldmann06,Znidaric09,Marshall17} and can autonomously generate Bell-nonlocal steady states when only Bell-local states were previously possible.

\section{Conclusion}
This paper has analysed a simple two-qubit thermal machine and found that Bell-nonlocal steady states are possible, without having to introduce additional resources. This result circumvents the no-go theorem presented in \cite{Brask2022}, which assumed that the system Hamiltonian includes a time-independent free-evolution term. There are models, however, where this term is absent in theory and may only appear as a stochastic perturbation in practice.

When the machine operates perfectly, a strong symmetry can enable the generation of Bell nonlocality at arbitrarily high temperatures. This suggests that further exploring a general link between the algebraic properties of thermal machines and quantum correlations could be fruitful for optimising their design. 

Experimental imperfections were then introduced via a stochastic perturbation to the system Hamiltonian, resulting in local and collective dephasing noise. It was shown that if the sole source of noise is local noise, then the steady state is always Bell local. Introducing collective noise alongside the local noise boosts the maximum attainable CHSH value, and makes steady-state Bell nonlocality possible. This suggests that noise might improve the quantum correlations observed in previously-proposed thermal machines.

A further question is whether the degree of steady-state Bell nonlocality in the thermal machine analysed here can be improved with additional resources. While all of the steady states found here are diagonal in the Bell basis, and therefore could not be improved by single-copy local filtering \cite{Verstraete02}, multi-copy local filtering could provide an advantage. The model can also be extended to multi-qubit systems, raising the question of whether the multi-qubit machine can generate multi-partite steady-state Bell nonlocality.

\begin{acknowledgements}
I would like to thank Charles Downing and Jonatan Bohr Brask for useful discussions on an early draft of the paper. This work was supported via a Royal Society University Research Fellowship (URF/R1/201158).
\end{acknowledgements}

\bibliographystyle{quantum}
\bibliography{references}

\appendix
\section{System eigenstates and jump operators \label{app:eig and jump}}
The system Hamiltonian $\hat{H}_S$ has eigenstates  $|11\ra,|00\ra,|\psi_-\ra,|\psi_+\ra$ with corresponding eigenvalues $\lambda = 0,0,g,-g$. The jump operators $\hat{A}_\beta(\omega)$ with Bohr frequency $\omega$ are defined as
\begin{equation}
    \hat{A}_\beta(\omega) = \sum_{\lambda_j-\lambda_i = \omega} \la \lambda_i |\hat{\sigma}_{x\beta} |\lambda_j\ra |\lambda_i\ra\la \lambda_j|,
\end{equation}
where $|\lambda_j\ra$ is the eigenstate corresponding to eigenvalue $\lambda_j$.

\section{Horodecki criterion for the thermal state \label{app:thermal}}
In the Bell basis $\{|\phi_+\ra,|\psi_+\ra,|\phi_-\ra,|\psi_-\ra\}$ the thermal state \eqref{eqn:thermal state} is represented by the matrix
\begin{equation}
    \rho_\textrm{th} = \frac{\sech^2(\frac{g}{2T})}{4}\textrm{diag}\left[1,\ue^{g/T},1,\ue^{-g/T}\right].
\end{equation}
Plugging this into the Horodecki criterion yields the maximal CHSH value
\begin{equation}
    \label{eqn:chsh thermal state}
    \mathcal{B} = 2\sqrt{2}\tanh\left(\frac{g}{2T}\right).
\end{equation}
The thermal state is Bell nonlocal ($\mathcal{B} > 2$) below the critical temperature $T = g/(2\ln[\sqrt{2}+1])$.

\section{A temperature gradient cannot enhance the steady-state Bell nonlocality  \label{app:no boost}}
Consider two baths at temperatures $T_1 > T_2$. Start from the Bell parameter \eqref{eqn:chsh 1} for the steady states of the thermal machine and focus on the first term in the square root. Substituting in expressions for $\gamma_\beta(\omega)$ \eqref{eqn:gamma rates} leads to
\begin{equation}
\label{eqn:gam1 adx}
\frac{\Gamma_1 - \Gamma_2}{\Gamma_1+\Gamma_2} = \frac{[1-(1+\xi_1)p_1]+x[1-(1+\xi_2)p_2]}{[1+(1-\xi_1)p_1]+x[1+(1-\xi_2)p_2]},
\end{equation}
where $x = J_2(g)/J_1(g)$ and for brevity the explicit frequency dependence of the distributions $p_\beta$ has been omitted in the notation.

The cases of two spin baths, a spin bath and a bosonic bath, and two bosonic baths, need to be considered separately. As the derivations are all similar, only the calculation for two spin baths will be presented. In this case, $\xi_1=1=\xi_2$ and \eqref{eqn:gam1 adx} takes the form
\begin{align}
    \frac{\Gamma_1 - \Gamma_2}{\Gamma_1+\Gamma_2} &= \frac{(1-2p_1)+x(1-2p_2)}{1+x},\nonumber\\
    &= \frac{\tanh\left(\frac{g}{2T_1}\right)+x\tanh\left(\frac{g}{2T_2}\right)}{1+x}
\end{align}
where the second line substituted in the Fermi-Dirac distribution (for the two spin baths). The hyperbolic tangent at the high temperature is smaller than at the low temperature so that the right-hand side is less than $\tanh(g/2T_2)$. Comparing this result to the Bell parameter for equal temperature baths \eqref{eqn:chsh thermal} immediately implies $\mathcal{B}(T_1,T_2) < \mathcal{B}(T_2,T_2)$.

\section{Steady state when noise is present \label{app:noisy results}}
When local and collective noise is included the steady state is unique,
\begin{align}
    \label{eqn:rho steady noise}
    \hat{\rho}_\infty &= (a+d)|\phi_+\rangle\langle \phi_+| + (b+c)|\psi_+\rangle\langle \psi_+| \nonumber \\ &+ (a-d)|\phi_-\rangle\langle \phi_-| + (b-c)|\psi_-\rangle\langle \psi_-|.
\end{align}
The coefficients are given by
\begin{align}
a &= \frac{1}{4} - \frac{\gamma_c}{4N}\frac{(\Gamma_1-\Gamma_2)^2}{\Gamma_1+\Gamma_2},\\
b &= \frac{1}{4} + \frac{\gamma_c}{4N}\frac{(\Gamma_1-\Gamma_2)^2}{\Gamma_1+\Gamma_2},\\
c &= \frac{(2\gamma_c+\gamma_l)}{4N}(\Gamma_1 - \Gamma_2),\\
d &= -\frac{\gamma_l}{4N}(\Gamma_1-\Gamma_2),
\end{align}
where $\Gamma_1$ and $\Gamma_2$ were defined in \eqref{eqn:big gammas}, $\gamma_l = (\gamma_{l_1}+\gamma_{l_2})/2$ and $N = 4\gamma_l^2+8\gamma_c\gamma_l + (\gamma_c+\gamma_l)(\Gamma_1+\Gamma_2)$.

If $T$ is the $3\times 3$ spin-correlation matrix (see the discussion preceding \eqref{eqn:chsh 1}), then for steady states of the form \eqref{eqn:rho steady noise} the eigenvalues of $T^\textrm{T} T$ are
\begin{equation}
    \label{eqn:TT eigs}
    \lambda_1 = 4(c+d)^2, \quad \lambda_2 = 4(c-d)^2, \quad \lambda_3 = 4(a-b)^2.
\end{equation}
The Bell parameter presented in the main text \eqref{eqn:horodecki noise} is found by  substituting  the expressions for $a,b,c$ and $d$ into \eqref{eqn:TT eigs}, and observing that the two largest eigenvalues are $\lambda_1$ and $\lambda_2$. 

\section{Conditions for steady-state Bell-nonlocality when noise is present \label{app:noisy nl conds}}
In this appendix the conditions for a Bell-nonlocal steady state are derived. To this end, fix $\gamma_l$ and express the Bell parameter \eqref{eqn:horodecki noise} in terms of the ratio $R = \gamma_l/\gamma_c$,
\begin{equation}
    \mathcal{B} = \frac{2\sqrt{R^2+2R+2}(\Gamma_1-\Gamma_2)}{(4\gamma_L+\Gamma_1+\Gamma_2)R + (8\gamma_l+\Gamma_1+\Gamma_2)}.
\end{equation}
Applying the requirement that $\mathcal{B}>2$ yields a quadratic inequality in $R$. The possible values of $R$ for which a CHSH inequality is violated are in the range $R_- < R < R_+$, where $R_\pm$ are the roots of the quadratic. The root $R_-$ is negative, and $R$ must be positive, so the only relevant root is
\begin{align}
    R_+ &= \bigg[\frac{1}{2}(\Gamma_1-\Gamma_2)\sqrt{\Gamma_1\Gamma_2+2\gamma_l(\Gamma_1+\Gamma_2)+8\gamma_l^2}\nonumber\\ &-(\Gamma_1\Gamma_2+3\gamma_l(\Gamma_1+\Gamma_2)+8\gamma_l^2)\bigg]\frac{1}{D},
\end{align}
where $D = \Gamma_1\Gamma_2+2\gamma_l(\Gamma_1+\Gamma_2)+4\gamma_l^2$. Imposing the constraint that this root is also positive leads to the condition
\begin{equation}
    (\Gamma_1+\Gamma_2-8\gamma_l)^2-8\Gamma_1\Gamma_2-128\gamma_l^2>0.
\end{equation}
This quadratic inequality is satisfied for $\Gamma_1>(3+2\sqrt{2})\Gamma_2+ 8(1+\sqrt{2})\gamma_l$ and $\Gamma_2>(3+2\sqrt{2})\Gamma_1+ 8(1+\sqrt{2})\gamma_l$. However, $\Gamma_1 > \Gamma_2$ so the second inequality never holds. The CHSH inequality is therefore violated when $\Gamma_1>(3+2\sqrt{2})\Gamma_2+ 8(1+\sqrt{2})\gamma_l$ and $R < R_+$. Recall that $R$ also needs to be consistent with the approximations involved in deriving the master equation (see the discussion under \eqref{eqn:liouvillian perturbed}).

\end{document}